\documentclass[twocolumn,showpacs,preprintnumbers,amsmath,amssymb]{revtex4}
\usepackage{epsfig}
\usepackage{graphicx}
\usepackage{dcolumn}

\newcommand{\beq}{\begin{equation}}
\newcommand{\eeq}{\end{equation}}
\usepackage{graphicx}
\usepackage{amssymb}

\newcommand{\beqa}{\begin{eqnarray}}
\newcommand{\eeqa}{\end{eqnarray}}

 \newcommand{\om}{\omega} 
 
\newcommand{\ra}{\rangle}

\def\pra#1{{ Phys.\ Rev. A\/} {\bf#1}} \def\prb#1{{ Phys.\ Rev. B\/} {\bf#1}}
 \def\prl#1{{ Phys.\ Rev.\
Lett.} {\bf#1}}

\begin{document}



\title{Sudden Death of Entanglement of Two Jaynes-Cummings Atoms  }

\author{Muhammed  Y\"{o}na\c{c}}

\author{Ting \  Yu}
\email{ting@pas.rochester.edu}

\author{J.\ H.\   Eberly}

\affiliation{ Rochester Theory Center for Optical Science and
Engineering, and  Department of Physics and Astronomy, University
of Rochester, New York 14627, USA }


\date{April 6, 2006}

\begin{abstract}

We investigate entanglement dynamics of two isolated atoms, each in its own Jaynes-Cummings cavity. We show analytically that initial entanglement has an interesting subsequent time evolution, including the so-called sudden death effect.

\end{abstract}


\maketitle

Entanglement is a defining feature of quantum
mechanics that makes fundamental distinctions  between quantum  and classical physics. As an unambiguous and quantifiable property of sufficiently simple multi-party quantum systems, entanglement has a definite time evolution that has recently begun to be studied in several contexts \cite{Yu-Eberly0203, dio, prlye, hal,  optye, privman, huetal, phys}. 

Entanglement in a quantum system may deteriorate due to interaction with background noise or with other systems usually 
called environments.  Interest was originally concerned with
the consequences for quantum measurement and the quantum-classical transition \cite{Zur, Zeh, cal}. More recently, entanglement decoherence  has been studied in connection with obstacles to realizing various quantum information processing schemes.
Particularly,  we have  shown
that entanglement can decay to zero abruptly, in a finite
time, a phenomenon termed entanglement sudden death \cite{prlye,optye}.

\begin{figure}[!b]
\epsfig{file=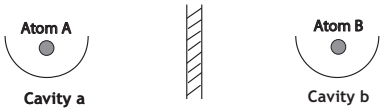, width=5.0 cm}
\caption{{\footnotesize \label{fig0}  This is a schematic diagram of the double J-C model used in this paper.
There is no communication between the cavities.}}
\end{figure}

The purpose of this paper is to examine two interesting time-evolving quantum systems that have no route for mutual iteraction, but whose mutual entanglement nevertheless evolves in an unusual way. We have chosen the ``double Jaynes-Cummings" model consisting of two two-level atoms. Each one is in a perfect one-mode near-resonant cavity and interacts with its initially unexcited cavity mode, but each is completely isolated from the other atom and cavity. 

By tracing over the cavity modes at time $t$ we are left with a mixed state of atoms $A$ and $B$ similar to that treated previously by us \cite{prlye}, but in this case the underlying intra-cavity dynamics are quite different since the cavities are here treated as lossless rather than as perfect reservoirs. The evolution of the entanglement of the non-interacting atoms here shows striking new features. 

By tracing cavity modes we are forcibly creating a two-qubit scenario, and various measures of entanglement are available. For a pair of qubits, all of them are equivalent, in the sense that when any of them indicates no entanglement (separable states), the others \cite{measures} also indicate no entanglement. Throughout the paper we will use Wootters' \cite{Wootters} concurrence $C(\rho)$ as the conveniently normalized entanglement measure ($1 \ge C \ge 0$).

The double Jaynes-Cummings  Hamiltonian for our system  may be written as,
\beqa
H_{\rm tot}&=&\om\sigma_z^A + \om\sigma_z^B + g(a^{\dagger}\sigma_{-}^A +
 a\sigma_{+}^A) \nonumber\\
 && + g(b^{\dagger}\sigma_{-}^B +
 b\sigma_{+}^B) + \nu a^{\dagger}a + \nu b^{\dagger}b.
\label{total}
\eeqa
Clearly there will be no interaction between atom $A$
and atom $B$ or between cavity $a$ and cavity $b$. The eigenstates of this Hamiltonian are products of the dressed states of the separate JC systems, which are well known \cite{JC}.

For greatest simplicity, we assume that both cavities are prepared initially in the vacuum state $|0_a\rangle\otimes |0_b\rangle$ and the two atoms are in a pure entangled state specified below. Under 
these assumptions,  there is never more than one photon in each
cavity, so the cavity mode is essentially equivalent to a two-level system.  This allows a uniform measure of quantum entanglement -- concurrence -- for both atoms and the cavity modes. 

In that connection we note that there are, in principle, six different concurrences that provide information about the overall entanglements that may arise. With an obvious notation we can denote these as $C^{AB}, ~C^{ab}, ~C^{Aa}, ~C^{Bb}, ~C^{Ab}, ~C^{Ba}$. Symmetry considerations can provide natural relations among these, which we will report elsewhere \cite{future}. Here we confine our attention to $C^{AB}$.

For a partially entangled atomic pure state that is a combination of the Bell states usually denoted $|\Psi^{\pm}\ra$, we have
\beq  
|\Psi_{\rm atom}\rangle=\cos \alpha|\uparrow\downarrow\rangle +
\sin\alpha|\downarrow\uparrow\rangle 
\eeq
with the first index  denoting the state of atom $A$ and
the second denoting the state of atom $B$ ($ \uparrow=$ excited state $\downarrow=$ground state),  
the initial state for the total system (\ref{total}) is given by
\begin{eqnarray}
|\Psi_0\rangle&=&(\cos\alpha|\uparrow\downarrow\rangle +
\sin\alpha |\downarrow\uparrow\rangle)\otimes |00\rangle \nonumber \\
&=&\cos\alpha|\uparrow\downarrow00\rangle +
\sin\alpha|\downarrow\uparrow00\rangle.
\end{eqnarray}
Then the solution of the model in terms of the  standard basis can be written as,
\beqa
|\Psi(t)\rangle&=&x_1|\uparrow\downarrow00\rangle +
x_2|\downarrow\uparrow00\rangle \nonumber\\
&&+ x_3|\downarrow\downarrow10\rangle
+ x_4|\downarrow\downarrow01\rangle,
\label{final4}
\eeqa
where the coefficients stand for the following time-dependent formulas:
\begin{eqnarray}
\label{coef}
x_1&=&(Le^{-i\lambda^+ t} + Me^{-i\lambda^-
t})\cos\alpha\nonumber\\
x_2&=&(Le^{-i\lambda^+ t} + Me^{-i\lambda^- t})\sin\alpha \nonumber \\
x_3&=&N(e^{-i\lambda^+ t} - e^{-i\lambda^-
t})\cos\alpha\nonumber\\
 x_4&=&N(e^{-i\lambda^+ t} -
e^{-i\lambda^- t})\sin\alpha.
\end{eqnarray}
Note here  that $\lambda^-$ and $\lambda^+$ are given by
\beq
\lambda^{\pm}=\nu+\frac{\Delta}{2}\pm\frac{\sqrt{\Delta^2+G^2}}{2},
\eeq
where $\Delta=\omega-\nu$ is the detuning and $G=2g$
represents the strength of interaction between the atoms and
cavities. The constant coefficients $L, M,$ and $N$ are given by
\beqa
L&=&\frac{1}{2}\left(1+\frac{\Delta}{\sqrt{\Delta^2+G^2}}\right)\\
M&=& \frac{1}{2}\left(1-\frac{\Delta}{\sqrt{\Delta^2+G^2}} \right) \\
N&=&\frac{G}{2\sqrt{\Delta^2+G^2}}.
\eeqa

The information about the entanglement of two atoms is contained in the reduced density
matrix  $\rho^{AB}$  for the two atoms which can be obtained from  (\ref{final4}) by tracing out the photonic parts
of the total pure state. The explicit  $4\times4$ matrix written in the basis
$|\uparrow\uparrow\rangle,|\uparrow\downarrow\rangle,|\downarrow\uparrow\rangle,
|\downarrow\downarrow\rangle$ is given by
\begin{equation}
\rho^{AB}=\left(
  \begin{array}{cccc}
    0 & 0 & 0 & 0 \\
    0 & |x_1|^2 & x_1x^*_2 & 0 \\
    0 & x_1^*x_2 & |x_2|^2 & 0 \\
    0 & 0 & 0 & |x_3|^2+|x_4|^2 \\
  \end{array}
\right)
\label{density1}
\end{equation}
which is in the ``normal" or ``standard" form of two-qubit mixed state we have noted previously \cite{YuEberly-arX05}. The time-dependent matrix elements are given by (\ref{coef}).

\begin{figure}[!t]
\epsfig{file=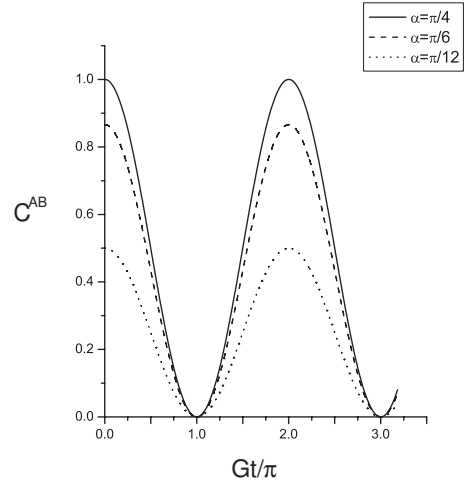, width=6.0 cm}
\caption{{\footnotesize \label{fig1}  The concurrence for atom-atom entanglement with
the initial atomic state $|\Psi_{\rm atom}\rangle=\cos \alpha|\uparrow\downarrow\rangle +
\sin\alpha|\downarrow\uparrow\rangle$  for zero detuning $\Delta=0$ }}\end{figure}

It can be shown that the concurrence of the density matrix (\ref{density1}) is given by
\begin{equation}
C^{AB}(t)=|\sin2\alpha|[1-4N^2\sin^2(\delta t/2)]
\end{equation}
where
\[\delta=\lambda^+ - \lambda^-=\sqrt{\Delta^2+G^2}.\]
A  particularly interesting example is the case of zero detuning ($\Delta=0$)  in which the concurrence becomes $C^{AB}(t)=|\sin2\alpha|\cos^2(Gt/2)$, which is shown in Fig. \ref{fig1}.

Alternatively, the initial state for the total system may be based on a combination of the other two Bell states $|\Phi^{\pm}\ra$: 
\begin{equation}
|\Phi_0\rangle=\cos\alpha |\uparrow\uparrow00\rangle+\sin\alpha |\downarrow\downarrow00\rangle,
\end{equation}
in which case the state of the total system at time $t$ can be expressed  in the standard basis:
\beqa
|\Phi(t)\rangle&=&x_1|\uparrow\uparrow00\rangle +
x_2|\downarrow\downarrow11\rangle\nonumber \\
&& + x_3|\uparrow\downarrow01\rangle
+x_4|\downarrow\uparrow10\rangle +x_5 |\downarrow\downarrow00\rangle \label{final3}
\eeqa
where the coefficients are now given by
\begin{eqnarray}
x_1&=&(Le^{-i\lambda^{+}t} + Me^{-i\lambda^{-}t})^2\cos\alpha \nonumber \\
x_2&=&LM(e^{-i\lambda^{+}t}-e^{-i\lambda^{-}t})^2\cos\alpha \nonumber \\
x_3&=&N(e^{-i\lambda^{+}t}-e^{-i\lambda^{-}t})
(Le^{-i\lambda^{+}t} + Me^{-i\lambda^{-}t})\cos\alpha\nonumber\\
x_4&=&N(e^{-i\lambda^{+}t}-e^{-i\lambda^{-}t})
(Le^{-i\lambda^{+}t} + Me^{-i\lambda^{-}t})\cos\alpha\nonumber \\
x_5&=&\sin\alpha.
\end{eqnarray}

In the basis of $|\uparrow\downarrow\rangle,
|\uparrow\uparrow\rangle, |\downarrow\downarrow\rangle,
|\downarrow\uparrow\rangle$ the reduced density matrix $\rho^{AB}$
is now found to be another example of ``standard" two-qubit mixed state:
\begin{equation}
\rho^{AB}=
\left(
  \begin{array}{cccc}
    |x_3|^2 & 0 & 0 & 0 \\
    0 & |x_1|^2 & x_1x_5 & 0 \\
    0 &  x_1^* x_5 &|x_2|+|x_5|^2 & 0 \\
    0 & 0 & 0 & |x_4|^2 \\
  \end{array} 
\right)
\end{equation}
and the concurrence for this matrix is given by
$C^{AB}(t)=\max\{0, f(t)\}$ 
where 
\beqa
f(t)&=&2|x_1||x_5|-2|x_3||x_4|\nonumber\\
&=& \big(1-4N^2\sin^2(\delta
t/2)\big)\big(|\sin2\alpha|\nonumber\\
&&-8N^2\sin^2(\delta t/2)\cos^2\alpha\big).
\eeqa
For $\Delta=0$, it becomes
\beq
f(t)=\cos^2(Gt/2)(|\sin2\alpha|-\sin^2(Gt/2)\cos^2\alpha).
\eeq

\begin{figure}
\begin{center}
  \includegraphics[width=6 cm]{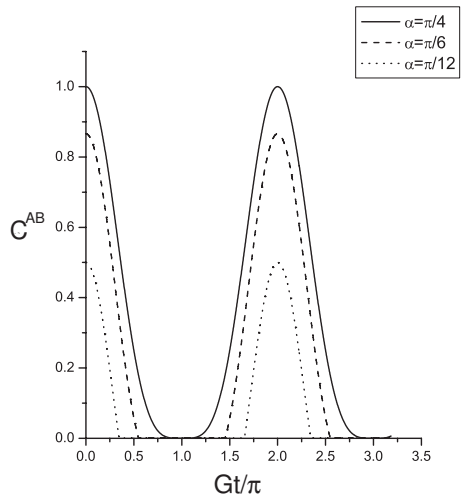}\\
  \caption{The concurrence for atom-atom entanglement with the initial atomic state 
 $|\Phi_{\rm atom}\rangle=\cos \alpha|\uparrow\uparrow\rangle +
\sin\alpha|\downarrow\downarrow\rangle$  for zero detuning $\Delta=0$.}\label{fig2}
  \end{center}
\end{figure}

Unlike the previous case,  Fig. \ref{fig2} shows that entanglement  can fall abruptly to zero (the two lower curves in the figure), and will remain zero for a period of
time before entanglement recovers.  The length of the time interval for the zero entanglement is dependent
on the degree of entanglement of the initial state.  The smaller the initial degree of entanglement is, the longer the state will stay in the disentangled separable state.

In summary,  let us emphasize again that there are 
no communications or interactions between the two atoms. The main result is the appearance of entanglement sudden death in a new environment, the first instance of sudden death without decoherence in the traditional sort. That is, because the cavities in our double Jaynes-Cummings model are lossless, they are as far from being standard decoherence reservoirs as possible. Nevertheless, we have shown that the non-interacting and non-communicating atoms $A$ and $B$ can abruptly lose their entanglement with each other. Given the lossless nature of the evolution, one can expect the resurrection of the original entanglement value to occur in a periodic way following each sudden death event, as is evident in Fig. \ref{fig2}.  Finally, we point out  that the onset of disappearance of two-atom entanglement is due to the information loss of atomic dynamics to the cavity-modes manifested by the operation of tracing over the cavity-variables.  On the other hand, it is the ``small"  numbers of the cavity-modes that lead to the  entanglement resurrection.  Namely, the lost information will come back to the atomic systems in finite times--a memory effect  that is often associated with the Rabi frequency in J-C cavities.  Clearly, the entanglement  of the two atoms will remain constant without their interactions  with the local cavities.

\section*{Acknowledgment}
We acknowledge interesting conversations with members of the University of Maryland - NIST Joint Quantum Institute. Financial support is provided by ARO via Grant W911NF-05-1-0543.


\begin{thebibliography}{99}

\bibitem{Yu-Eberly0203} T. Yu and J.H. Eberly, \prb{66}, 193306
(2002) and \prb {68}, 165322 (2003).

\bibitem{dio}L. Diosi, in {\em Irreversible Quantum Dynamics},
edited by F. Benatti and R. Floreanini (Springer, New York, 2003),
pp. 157-163.

\bibitem{hal} P.J. Dodd and J.J. Halliwell, \pra {69},
052105 (2004), and P.J. Dodd, \pra {69}, 052106 (2004).

\bibitem{prlye} T. Yu and J. H. Eberly, \prl {93}, 140404 (2004).

\bibitem{optye} T. Yu and J.H. Eberly,  Opt. Comm.  (in press,  2006).

\bibitem{privman} D. Tolkunov, V.   Privman, and P. K.  \  Aravind,
Phys. Rev. A {\bf 71}, 060308 (2005).

\bibitem{huetal}S. \ Shresta, C.\  Anastopoulos, \ A.\  Dragulescu, and B. L. Hu,
Phys. Rev. A  {\bf 71}, 022109 (2005).


\bibitem{phys} F.\  Mintert, A.\ R. \ R. \ Carvalho, M.\ Kus, A.   Buchleitner,
                         Phys. Rep.  {\bf  415}, 4,  207(2005). 

\bibitem{Zur} W. H. Zurek,  Phys.  Rev. D {\bf 24}, 1516 (1981); Phys. Today {\bf 44}, 36 (1991).

\bibitem{Zeh} E.  Joos, H. D. Zeh, C.  Kiefer, D.  Giulini,  K.  Kupsch, I. -O. Stamatescu,
{\it Decoherence and the Appearance of a Classical World in Quantum Theory} (Springer, Germany,  2003).

\bibitem{cal} A. Caldeira and A. Leggett, Phys. Rev. A 31, 1059 (1985).


\bibitem{measures} For a two-qubit state entanglement a set of alternative measures of entanglement equivalent to concurrence in the sense mentioned in the text includes these: entropy of formation, tangle, Schmidt number, and negativity.

\bibitem{Wootters} W.\ K. \ Wootters, \prl {80}, 2245 (1998).

\bibitem{JC} E.T. Jaynes and F.W. Cummings, Proc. IEEE, {\bf 51}, 89 (1963).

\bibitem{future} M.\ Y\"{o}nac, T.\  Yu, and J.\ H.\  Eberly,  in preparation (2006).

\bibitem{YuEberly-arX05} See T. Yu and J.H. Eberly, arXiv: quant-ph/0503089.

\end{thebibliography}
\end{document}